\documentclass[aps,prl,floatfix,amsfonts,amssymb,twocolumn,superscriptaddress,showpacs]{revtex4}
\usepackage[dvips]{color}
\usepackage[dvips]{graphics}
\usepackage{graphicx}
\def\cT{{\cal T}}
\def\vv{{\bf v}}
\def\vn{{\bf n}}
\def\vp{{\bf p}}
\def\vq{{\bf q}}
\def\vA{{\bf A}}
\def\vR{{\bf R}}
\newcommand{\F}{\mathfrak{F}}
\newcommand{\G}{\mathfrak{G}}
\newcommand{\hG}{\hat\mathfrak{G}}
\newcommand{\grad}{\mbox{\boldmath$\nabla$}}
\newcommand{\sgn}{\mbox{sgn}}
\renewcommand{\Im}{\mathfrak{Im}\,}
\begin{document}
\title{Nonlinear magnetic field dependence of the conductance in d-wave \\ NIS tunnel junctions}
\author{M.~Fogelstr\"om}
\affiliation{Department of Physics and Astronomy, Northwestern University, Evanston, IL 60208 USA}
\affiliation{Institute of Theoretical Physics, Chalmers University of
             Technology and G\"oteborgs University, S-41296 G\"oteborg, Sweden}
\author{D.~Rainer}
\affiliation{Department of Physics and Astronomy, Northwestern University, Evanston, IL 60208 USA}
\affiliation{Physikalisches Institut, Universit\"at Bayreuth, D-95440 Bayreuth, Germany}
\author{J.A.~Sauls}
\affiliation{Department of Physics and Astronomy, Northwestern University, Evanston, IL 60208 USA}
\email[{Email Correspondence to: }]{sauls@snowmass.phys.nwu.edu}
\preprint{\hspace*{-18cm}{\tiny Draft: \today}}
\begin{abstract}
The ab-plane NIS-tunnelling conductance in d-wave superconductors shows a zero-bias conductance peak which is
predicted to split in a magnetic field. In a pure d-wave superconductor the splitting is linear for fields
small on the scale of the thermodynamic critical field. The field dependence is shown to be nonlinear, even at
low fields, in the vicinity of a surface phase transition into a local time-reversal symmetry breaking state.
The field evolution of the conductance is sensitive to temperature, doping, and the symmetry of the
sub-dominant pairing channel.
\end{abstract}
 \maketitle

Tunnelling into an {\sl unconventional} superconductor probes the quasiparticle spectrum associated with a
surface superconducting state that is typically strongly deformed, and may have different local symmetry, than
the bulk order parameter (OP) \cite{lof01}. Indeed, the OP of a pure ${\text{d}_{x^2-y^2}}$ superconductor is
suppressed at a [110] surface. The suppression is associated with the formation of zero-energy Andreev states
that are bound to the surface (ABS). The ABS is observable as a zero-bias peak (ZBCP) in the tunnelling
conductance \cite{buc95b,tan95}. This property makes quasiparticle tunnelling in unconventional
superconductors a sensitive probe of broken symmetry. A key signature of a surface ABS is its dependence on an
applied magnetic field, or more precisely on screening currents flowing along the surface. For an external
magnetic field perpendicular to the conducting plane and parallel to the surface, screening currents produce a
Doppler shift the ABS \cite{fog97a} which is observable as a splitting, $2\delta(H)$, of the ZBCP with applied
field. Some of the predicted field dependence of the splitting has been verified experimentally by Covington
et al. \cite{cov97}, Aprili et al.\cite{apr99}, and Krupke and Deutscher \cite{kru99} for
YBa$_2$Cu$_3$O$_{7-p}$ (YBCO) near optimal doping.

It was shown theoretically that a surface phase transition to a state with spontaneously broken time-reversal
($\cT$) symmetry 
should occur for a d-wave superconductor with strong surface pair breaking if there is an attractive
subdominant pairing channel \cite{sig95,mat95,fog97a}. A signature of this transition in the tunnelling
conductance is a spontaneous splitting of the ZBCP that develops below a surface phase transition temperature,
$T_s$. Such a transition was observed by Covington et al. \cite{cov97} for \texttt{Cu|I|YBCO} junctions, and
also by Krupke, et al. \cite{kru99} and Dagan et al. \cite{dag01} for \texttt{In|I|YBCO} junctions. The latter
authors found that the spontaneous splitting of the ZBCP occurred only above a critical doping level, $p_c$,
close to optimal doping. They also showed that the magnitude of the field-induced splitting of the ZBCP is
suppressed at low fields for slightly underdoped materials ($p\lesssim p_c$), while for $p\thickapprox p_c$
$\delta(H)$ is nonlinear in $H$ and onsets rapidly. For slightly overdoped YBCO ($p\gtrsim p_c$) $\delta(H)$
increases linearly with $H$ at low field from a non-zero splitting in zero field.

\begin{figure}[ht]
\centerline{\scalebox{.41}{\includegraphics{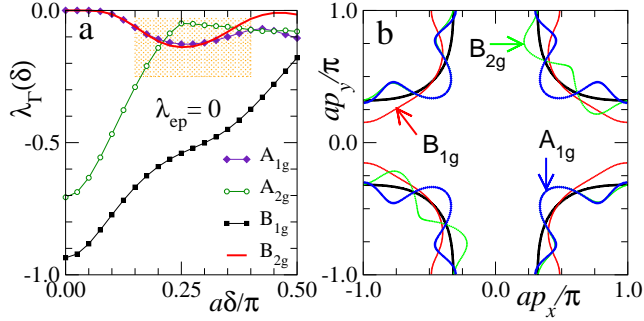}}} \caption[]{Eigenvalues and
eigenfunctions for the singlet, 1D irreducible representations of $D_{4h}$ based on Eq.
\ref{pairing_interaction}. (a) Variation of $\lambda_{\Gamma}$ with incommensurability parameter, $\delta$,
for $\xi_{\text{af}}=2a$ and $\lambda_{\text{ep}}=0$. (b) Basis functions for the dominant ($\rm{B}_{1g}$) and
the sub-dominant ($\rm{B}_{2g}$ and $\rm{A}_{1g}$) pairing channels for $a\delta=\frac{\pi}{4}$. The black
curve is the Fermi surface. } \label{EigenBasis}
\end{figure}

In this Letter we report calculations of the field dependence of the conductance of NIS tunnel junctions in
YBCO. We consider a pairing interaction that includes both magnetic and phonon-mediated pairing channels
\cite{rad92,buc95b,esc01},
\begin{equation}
\lambda(\vq)=\sum_{\delta_{x,y}}^{\pm \delta}
\frac{\lambda_{\text{af}}/4}{1+4\xi^2_{\text{af}}(\cos^2\frac{q_x-\delta_x}{2}+\cos^2\frac{q_y-\delta_y}{2})}-
\lambda_{\text{ep}} \,. \label{pairing_interaction}
\end{equation}
The repulsive magnetic contribution to $\lambda(\vq)$ results from exchange and short-range spin correlations,
and is assumed to dominate a weaker, attractive electron-phonon contribution to the pairing interaction. The
interaction depends on the correlation length, $\xi_{\text{af}}$, the incommensurate wave vectors for the
spin-excitation spectrum, $\delta_{x,y}$, and the electron-phonon and magnetic coupling strengths,
$\lambda_{\text{ep}}$ and $\lambda_{\text{af}}$. The quasiparticle dispersion relation is modelled by the
tight-binding parametrization for YBCO taken from Ref. \cite{rad94}. The phonon-mediated interaction is
assumed to be attractive in the spin-singlet, s-wave ($\text{A}_{\text{1g}}$) pairing channel. Indirect
evidence for an attractive sub-dominant s-wave interaction in the cuprates comes from ab-plane tunnelling
measurements on Pr-doped YBCO which show strongly suppressed transitions, $T_c \lesssim 20\,\text{K}$, and no
ZBCP indicative of d-wave superconductivity \cite{cov00}. For this pairing interaction and band structure we
calculated the eigenvalues $\lambda_\Gamma$ and eigenfunctions, $\eta_{\Gamma}(\vp_f)$, of the linearized gap
equation following Refs. \cite{buc95b,esc01}. The eigenfunctions determine the anisotropy of the pairing state
in momentum space and form basis functions for the irreducible representations ($\Gamma$) of the point group,
$D_{4h}$. The eigenvalues determine the instability temperatures $T^\Gamma_c$, for pairing into  states with
symmetry dictated by the corresponding irreducible representation. The pairing interaction in Eq.
\ref{pairing_interaction} allows us to examine the sensitivity of the dominant and sub-dominant pairing
channels to the parameters defining the model: $\xi_{\text{af}}$, $\delta_{x,y}$, and the coupling parameters,
$\lambda_{\text{ep,af}}$, and to consider the doping dependence via these parameters. We consider the
one-dimensional, spin-singlet representations, $\rm{A_{1g}},\rm{A_{2g}},\rm{B_{1g}},$ and $\rm{B_{2g}}$. The
most attractive eigenvalues calculated for the two-channel interaction are shown in Fig. \ref{EigenBasis}a as
a function of $\delta$. Note that all four channels have attractive eigenvalues for incommensurate
spin-fluctuations i.e. for $\vq$ vectors away from the $(\pm\pi/a,\pm\pi/a)$-points. This is a robust feature
of the model for $\xi_{\rm{af}}\lesssim 4\,a$. Secondly, the $\rm{B_{1g}}$ ($\text{d}_{x^2-y^2}$)
representation is the dominant pairing channel for $\delta\lesssim\pi/2a$. A third key point is that the
leading sub-dominant pairing channels are nearly degenerate over a wide range of values of the incommensurate
wavevector, $\delta$, for $\lambda_{\text{ep}}=0$. Thus, weaker interactions like the electron-phonon coupling
can play an important role in determining the sub-dominant pairing channel. For $\xi_{\rm{af}}=2a$ and
$a\delta_{x,y}=\frac{\pi}{4}$ varying the strength of $\lambda_{\text{ep}}$ allows us to tune the leading
sub-dominant pairing channel; increasing $\lambda_{\text{ep}}$ favors the $\rm{A_{1g}}$-channel, and it
becomes more attractive than the $\rm{B_{2g}}$ channel at $\lambda_{\text{ep}}\simeq 0.1\lambda_{\text{af}}$.

The eigenfunctions, $\eta_\Gamma(\vp_f)$, also depend on the parameters of the pairing model:
$\eta_{\rm{B_{1g}}}$, $\eta_{\rm{B_{2g}}}$ and $\eta_{\rm{A_{2g}}}$ depend most sensitively on
$\xi_{\text{af}}$ and $\delta_{x,y}$, while $\eta_{\rm{A_{1g}}}$ also depends on $\lambda_{\text{ep}}$. The
calculated basis functions are shown in Fig. \ref{EigenBasis}b for $\lambda_{\text{ep}}=0$;
$\eta_{\rm{B_{1g}}}$ is well described by the simplest $\text{d}_{x^2-y^2}$ basis function,
$\eta_{d}(\phi_p)=\sqrt{2}\cos(2 \phi_p)$, where $\phi_p$ is the angle $\vp_f$ makes with respect to the
$[100]$ direction of the crystal. For the $\rm{B_{2g}}$ channel the calculated basis function is not the
simplest ${\rm{d}_{xy}}$ basis function, $\eta_{d^\prime}(\phi_p)=\sqrt{2}\sin(2 \phi_p)$, but a higher
harmonic which to a good approximation is $\eta_{\rm{B_{2g}}}(\phi_p)\simeq 2\,|\sin(2 \phi_p)|\sin(6\phi_p)$.
The $\rm{A_{1g}}$ basis function is highly anisotropic for $\lambda_{\text{ep}}<0.1\lambda_{\text{af}}$ but
nearly isotropic for $\lambda_{\text{ep}}>0.1\lambda_{\text{af}}$. The $\rm{A_{1g}}$ basis function is to good
approximation given by, $\eta_{\rm{A_{1g}}}(\phi_p)=(2/\sqrt{1-2\alpha +5\alpha^2}) \big(\alpha
-(1-\alpha)\sin(2 \phi_p) \sin(6 \phi_p)\big)$, where the parameter $\alpha$ varies between $\alpha=0$ for
$\lambda_{\text{ep}}\ll 0.1\lambda_{\text{af}}$ and $\alpha=1$ for $\lambda_{\text{ep}}\gg
0.1\lambda_{\text{af}}$. In general $\alpha$ varies with doping and provides us with a convenient
parametrization of the effect of doping on the pairing eigenfunctions.

The basis functions, $\eta_\Gamma(\vp_f)$, $T_c$, and the relative coupling strengths of the sub-dominant
pairing channel are inputs to calculations of the surface phases and tunnelling conductance for d-wave models
of the cuprates. The surface OP, $\Delta(\vp_f,\vR)=\sum_{\Gamma}\Delta_{\Gamma}(\vR)\eta_\Gamma(\vp_f)$,
condensate momentum, $\vp_s$, and excitation spectrum are obtained, selfconsistently, from solutions for the
propagator, spectral density and OP describing the superconducting state. The propagator obeys the
quasiclassical transport equations \cite{buc95a}, and the OP, $\Delta(\vp_f,\vR)$, is determined by the
anomalous quasiclassical propagator, $\F(\vp_f,\vR;\epsilon_n)$, through the BCS gap equation,
\begin{equation}
\Delta(\vp_f,\vR)=T\sum_{\epsilon_n}\int d^2\vp_f^\prime\lambda(\vp_f-\vp_f')\,\F(\vp_f',\vR;\epsilon_n)\,.
\label{gapequation}
\end{equation}
The pairing interaction, $\lambda(\vp_f-\vp_f')$, is resolved in terms of its eigenfunctions and eigenvalues:
$\lambda(\vp_f-\vp_f')=\sum_{\Gamma}\lambda_{\Gamma}\eta_{\Gamma}(\vp_f)\eta_{\Gamma}^*(\vp_f')$, and we
eliminate the interaction parameters, $\lambda_{\Gamma}$, and cutoff frequency, $\epsilon_c$, in favor of the
measurable instability temperatures, $T^\Gamma_c$ \cite{buc95a}. The formulation and application of the
quasiclassical transport equations to surface states and tunnelling is described in Refs.
\cite{buc95a,buc95b,esc00}.

The surface superconducting phase is sensitive to magnetic fields that generate surface screening currents.
These currents give rise to Doppler shifts of the low-energy surface excitations, $\epsilon_{\text{\tiny
Doppler}}=\vv_f\cdot\vp_s(\vR)$, which modify the relative stability of surface phases described by different
local symmetries, and may induce surface phases with broken $\cal T$-symmetry. This type of {\sl
field-induced} transition leads to low-field nonlinearities in the field-dependence of the splitting of the
ZBCP, $\delta(H)$. The condensate momentum is given by the phase gradient and gauge-invariant coupling to the
vector potential, $\vp_s(\vR)=\frac{\hbar}{2}(\grad\vartheta-(e/c)\vA(\vR))$. For magnetic fields applied
along the crystal $\hat c$-axis, $\vp_s(\vR)$ is perpendicular both to $\hat c$ and to the surface normal
$\vn$, and of order $v_fp_s\simeq\Delta(H/H_o)$ where $H_o=(c/e)\Delta/v_f\lambda$ is the pairbreaking field
of order of a few Tesla in the cuprates \cite{fog97a}. The Fermi velocity, $v_f$ and density of states, $N_f$,
determine the zero-temperature penetration depth $1/\lambda^2=(4\pi e^2/c^2)\,N_f v_f^2$ in the clean limit,
and we take $\lambda/\xi_o=100$ for YBCO.

The conductance of a NIS tunnel junction measures an angle-average of the quasiparticle density of states
(DOS) at the interface. In the limit of a low-transmission tunnel barrier $({\cal D}\ll 1)$ the conductance is
given to leading order in $\cal D$ by the standard tunnelling result, $dI/dV=(1/R_N)\int_{>}\,d^2\vp_f
{\cal{D}}(\vp_f)N(\vp_f,\vR_s;{\rm{eV}})$ in the limit $T\ll T_c$, where the integral is over trajectories
with $\vp_f\!\cdot\!\vn > 0$, $R_N$ is the interface resistance,
$N(\vp_f,\vR_s;{\rm{eV}})=-\frac{1}{\pi}\Im\G^R(\vp_f,\vR_s;\epsilon\!=\!eV)$ is the angle-resolved local DOS
of the superconductor evaluated at the surface, $\vR_s$ and $\G^R$ is the diagonal element of the retarded
propagator, $\hG^R=\hG(i\epsilon_n\rightarrow\epsilon+i0^+)$. The angle-resolved local DOS is folded with the
trajectory dependent barrier transmission probability, ${\cal{D}}(\vp_f)$. The tunnelling barrier is modelled
by a transmission probability that is maximum for quasiparticles with Fermi momenta parallel to the surface
normal ($\vp_f||\vn$) and decreases with angle as ${\cal{D}}={\cal{D}}_0\,{\mbox{e}}^{-(\phi/\phi_c)^2}$ where
$\phi=\cos^{-1}(\hat{\vp}_f\cdot\vn)$ and $\phi_c$ is a measure of active tunnelling trajectories.
Qualitatively, thick junctions have narrow tunnelling cones, $\phi_c\ll\pi/2$, while thinner junctions
correspond to a wider range of active tunnelling trajectories.

\begin{figure}[ht]
\centerline{\scalebox{0.34}{\includegraphics{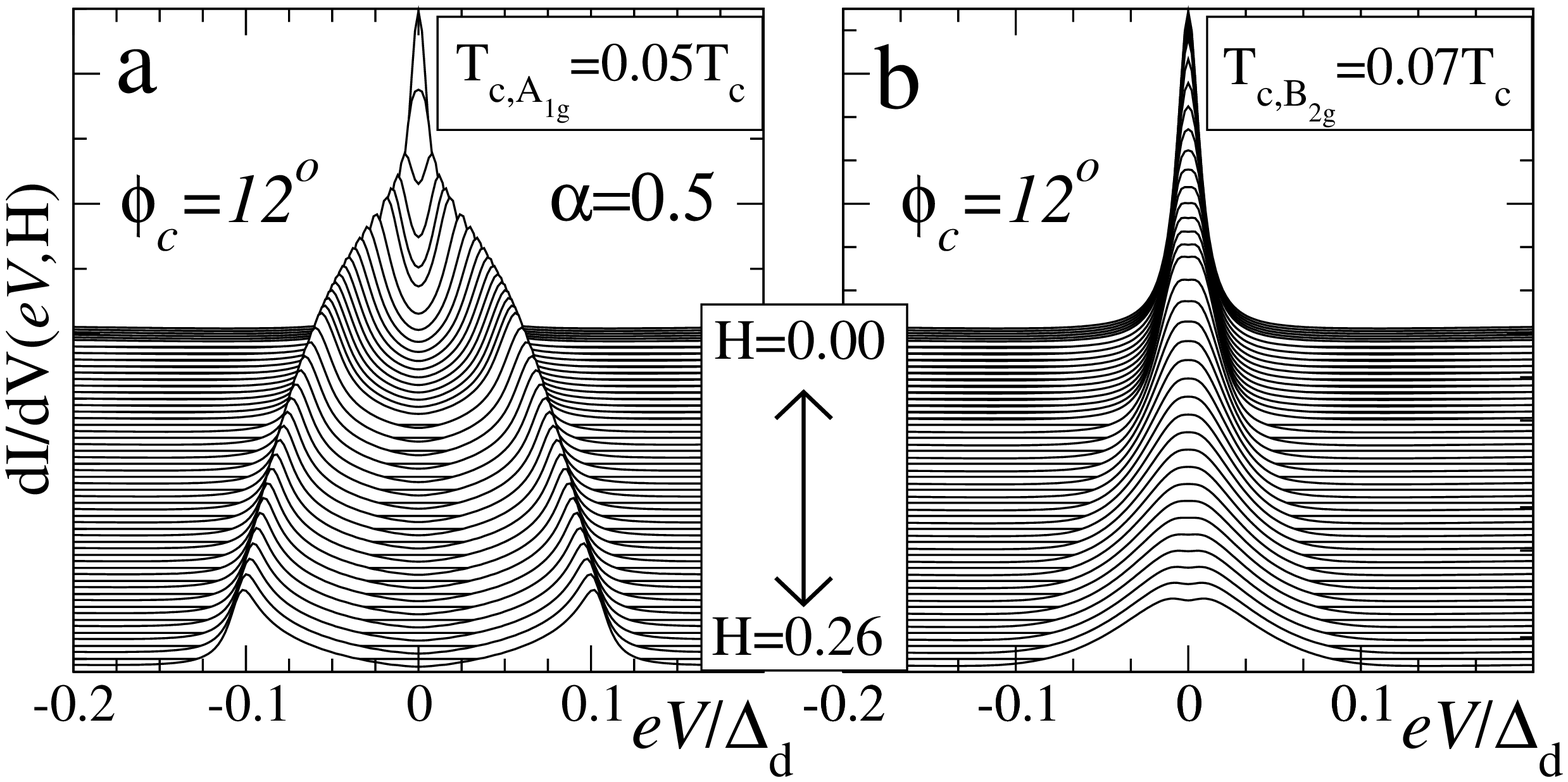}}} \caption[]{The field dependence of the
low-bias tunnelling conductance, $dI/dV$, calculated at different fields for a $\rm{d_{x^2-y^2}}$-wave
superconductor in the presence of a sub-dominant $\rm{A_{1g}}(\alpha=0.5)$ pairing interaction [panel a], and
a sub-dominant ${\rm{B_{2g}}}$ pairing interaction [panel b]. The pairing strength is too weak for a local
${\cal{T}}$-breaking state to form spontaneously at the surface.} \label{Conductances}
\end{figure}

In Fig. \ref{Conductances} we show results for the tunnelling conductances for the sub-dominant pairing
channels with $\rm{A_{1g}}$ and $\rm{B_{2g}}$ symmetries. The surface DOS was calculated at fixed temperature
just above a surface phase transition with broken $\cal T$ symmetry, $T\gtrsim T_s^\Gamma$. The conductances
were calculated in the clean limit at $T=0.1\,T_c$. Note the nonlinear field evolution shown in Fig.
\ref{Conductances}a for the $\rm{A_{1g}}(\alpha\!=\!0.5)$ channel. For the  $\rm{B_{2g}}$ channel the
splitting of the ZBCP is suppressed at low fields.

\begin{figure}[hbt]
\centerline{\scalebox{0.47}{\includegraphics{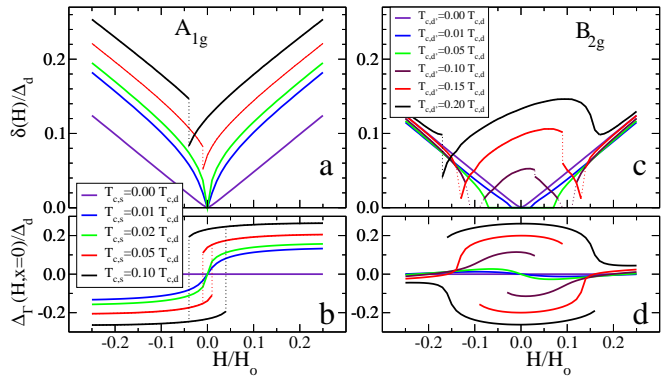}}} \caption[]{Splitting of the ZBCP, $\delta(H)$,
as a function of magnetic field and strength of the sub-dominant pairing channel, for $\phi_c=\pi/4$.
Field-dependence of the sub-dominant order parameter at the surface. Sub-dominant channels: (a,b)
$\rm{A_{1g}}$, (c,d) $\rm{B_{2g}}$. }\label{field_dep}
\end{figure}

The field induced splitting of the ZBCP, $\delta(H)$, is linear in $H$ for pure d-wave pairing \cite{fog97a},
as shown in Fig. \ref{field_dep}. The proximity to a broken $\cal T$-symmetry phase of the form
$\rm{d}+i\rm{s}$ enhances the splitting of the ZBCP and generates a field dependence of the shift in the
conductance peak that is nonlinear in $H$ for $T\gtrsim T_s$ (Figs. \ref{Conductances}a and \ref{field_dep}a).
By contrast, the field splitting for the $\rm{d_{xy}}$ channel is suppressed at low fields; but onsets above a
critical field of order $H^*_{\rm{d^\prime}}=(T_c^{\text{B}_{2g}}/T_c)H_o$ (Fig. \ref{field_dep}c). The field
dependence of the splitting, $\delta(H)$, and its sensitivity to a sub-dominant order parameter,
$\Delta_{\Gamma}$, can be understood by considering the ABS spectrum for an OP that is constant in space, but
breaks $\cal T$-symmetry, e.g. $\Delta(\vp_f)=\Delta_d(\vp_f)+i\Delta_\Gamma(\vp_f)$. The $\rm{d_{x^2-y^2}}$
component, $\Delta_{\text{d}}$, changes sign along the scattered trajectory, while the sub-dominant component
is unchanged.  Thus, $\Delta(\vp_f^\prime)^*=-\Delta(\vp_f)$ on each trajectory. From the retarded propagator,
$\G^R(\vp_f,\epsilon)$, we obtain the ABS pole at an energy shifted away from the Fermi level,
$\varepsilon_b(\vp_f)=-s_\Gamma\,\Delta_{\Gamma}(\vp_f)-\vv_f\cdot\vp_s$,
with $s_\Gamma=\sgn[\Delta_{\text{d}}(\vp_f)\Delta_{\Gamma}(\vp_f)]$.

\noindent In the absence of the Doppler shift, a sub-dominant OP shifts the ABS. The magnitude and sign of the
shift depend on the magnitude of the sub-dominant OP and the relative phase between $\Delta_{\text{d}}$ and
$\Delta_{\Gamma}$, $s_\Gamma=\pm$. The bound-state energies for trajectories, $\vp_f$ and $-\vp_f$, have
opposite signs; this splitting for time-reversed states results in spontaneous surface currents carried by the
bound states \cite{mat95,fog97a}. The two possible directions for the spontaneous current reflect the two-fold
degeneracy of the surface phase, i.e. $\Delta_d(\vp_f)\pm i\Delta_\Gamma(\vp_f)$. In a magnetic field
screening currents generate a Doppler shift of the surface bound states. For the sub-dominant $s$-wave OP the
Doppler term may generate a shift of the same sign as the spontaneous shift, or the reverse depending on the
polarity of the field and the relative phase of the subdominant OP. If the shifts are opposite in sign, the
bound state energies move toward the Fermi level with increasing field, \underline{until} it becomes
energetically favorable for the relative phase and bound state current to reverse sign. This reversal occurs
at a low field of order $H^*=(T_s/T_c)H_o$ for the sub-dominant s-wave channel as shown in Fig.
\ref{field_dep}b. Also shown is the field-induced sub-dominant OP that onsets rapidly with field for $T>T_s$,
and which is the source of the nonlinearity in $\delta(H)$ shown in Fig. \ref{field_dep}a. For  $T<T_s$ this
splitting is asymmetric about $H=0$, and hysteretic on a field scale of order $H^*$.

In the doping range where the leading sub-dominant channel is $\rm{B_{2g}}$, the field evolution is more
complex. The angular dependence of the $\rm{B_{2g}}$ OP, $\Delta_{\rm{d_{xy}}}\sim\sin
2(\phi_p-\frac{\pi}{4})$, implies that the bound state disperses through the Fermi level as a function of the
trajectory angle, $\phi=\cos^{-1}(\hat{\vp}_f\cdot\vn)$. For a $[110]$ surface half of the incident
trajectories correspond to ABS above the Fermi level and the other half have their energies below the Fermi
level. When added to their time-reversed partners, these states carry oppositely directed currents.
Field-induction of the $\rm{d_{xy}}$ OP for $T>T_s$ produces a different behavior than predicted for $s$-wave
sub-dominant pairing; the magnitude of $\Delta_{\rm{d_{xy}}}$ remains small, and the corresponding bound state
splittings are compensated by the Doppler term, until a threshold field is reached, again of order $H^*$,
beyond which the Doppler term cannot be compensated by the shift from the field-induced sub-dominant OP (see
Fig. \ref{field_dep}c-d). For $T<T_s$ the subdominant OP produces splitting of the ZBCP for $H=0$. The
low-field evolution is again different than that of the $s$-wave case; $\delta(H)$ decreases with $H$ until a
critical field of order $H^*$ is achieved. This field corresponds to the counter-moving branch of ABS
dispersing through the Fermi level. Further increasing the field can drive the sub-dominant OP to very small
values, in which case the splitting is given by the Doppler splitting and linear for $H\ll H_o$.

\begin{figure}[ht]
\centerline{\scalebox{0.30}{\includegraphics{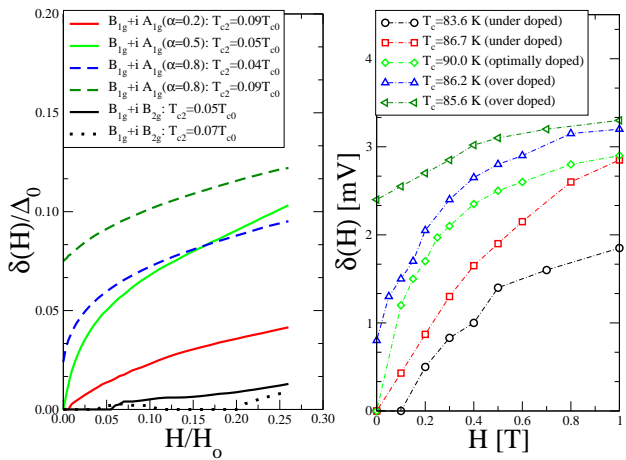}}} \caption[]{(a) Calculated field
dependence of the conductance maximum as a function applied field for different values of doping parameter
$\alpha$. The model for the transmission probability is ${\cal{D}}(\phi)={\cal{D}}_0 \exp[-(\phi/\phi_c)^2]$,
with $\phi_c=12^o$. (b) Measured field dependence of the conductance maximum as a function of applied field
for different values of doping as reported in Fig. 3 of Ref. \cite{dag01}. } \label{Fit}
\end{figure}

Many of these features appear to be observed in YBCO tunnel junctions. We conclude with a comparison of the
theoretical predictions of the surface ABS model for the tunnelling conductance, including the qualitative
predictions of the two-channel pairing model, with the tunnelling measurements reported in Ref. \cite{dag01}
for the field dependence of the tunnelling conductance for YBCO. Calculated splittings of the ZBCP,
$\delta(H)$, are shown in Fig. \ref{Fit}a, and are compared with the experimental results shown in Fig.
\ref{Fit}b. Experimental observations of the field dependence of the conductance peak in underdoped films are
in agreement with theoretical results based on a field-induced $\rm{d_{x^2-y^2}}+i\rm{d_{xy}}$ surface state;
there is a low-field threshold before a splitting appears, followed by an approximately linear increase in the
splitting with $H$. At optimal doping the linear field dependence is recovered with no spontaneous splitting.
Further increase in the doping level shows the nonlinear regime for slightly overdoped samples, which is
accounted for by a attractive s-wave subdominant pairing channel. The evolution with field and doping is
systematically reproduced by a ${\rm{B_{1g}}}+i {\rm{A_{1g}}}(\alpha)$ surface state with anisotropic
${\rm{A_{1g}}}$ component closest to optimal doping which evolves as the doping is increased to a nearly
isotropic ${\rm{A_{1g}}}$ sub-dominant pairing state. The evolution of the ZBCP indicates that the pairing
interaction changes with doping; the dominant pairing channels is ${\rm{B_{1g}}}$, with a sub-dominant
${\rm{B_{2g}}}$ component in the slightly underdoped regime which is overtaken by a sub-dominant
${\rm{A_{1g}}}$ component in the overdoped regime. The cross-over occurs close to optimal doping.

This work was supported by the NSF grant DMR 9972087, and the Swedish Research Council, VR.

\end{document}